# Edge states at nematic domain walls in FeSe films

Yonghao Yuan,[†] Wei Li,[*†‡] Bin Liu,[§] Peng Deng,[†] Zhilin Xu,[†] Xi Chen,[†‡] Canli Song,[†‡] Lili Wang,[†‡] Ke He,[†‡] Gang Xu,[*§] Xucun Ma,[†‡] and Qi-Kun Xue[*†‡]

[†]*State Key Laboratory of Low-Dimensional Quantum Physics, Department of Physics, Tsinghua University, Beijing 100084, China*

[‡]*Collaborative Innovation Center of Quantum Matter, Beijing 100084, China*

[§]*Wuhan National High Magnetic Field Center & School of Physics, Huazhong University of Science and Technology, Wuhan 430074, China*

[*]Wei Li Email: weili83@tsinghua.edu.cn Phone: +8610-62795838
[*]Gang Xu Email: gangxu@hust.edu.cn Phone: +8627-87792334-8120
[*]Qi-Kun Xue Email: qkxue@mail.tsinghua.edu.cn Phone: +8610-62795618

**ABSTRACT:**

Quantum spin Hall (QSH) effect is an intriguing phenomenon arising from the helical edge states in two-dimensional topological insulators. We use molecular beam epitaxy (MBE) to prepare FeSe films with atomically sharp nematic domain boundaries, where tensile strains, nematicity suppression and topological band inversion are simultaneously achieved. Using scanning tunneling microscopy (STM), we observe edge states at the Fermi level that spatially distribute as two distinct strips in the vicinity of the domain boundaries. At the endpoint of the boundaries, a bound state at the Fermi level is further observed. The topological origin of the edge states is supported by density functional theory calculations. Our findings not only demonstrate a candidate for QSH states, but also provide a new pathway to realize topological superconductivity in a single-component film.

FeSe is an intriguing material that can realize superconductivity enhancement,[1-4] topological non-trivial edge states[5, 6] and topological superconductivity.[7-12] Chemical potential and lattice strain are two main factors to manipulate the electronic properties and induce those emergent phenomena.

Nematicity,[13] defined as rotational symmetry breaking of electronic structures in iron-based superconductors, is an interesting phenomenon that emerges at low temperature. It manifests as electronic anisotropy observed in experiments.[14-20] For most of the iron-based superconductors, superconductivity arises with the suppression of nematic phase via carrier doping. Nematicity is also suppressed at the boundaries of nematic domains (domain walls), where nematicity is broken due to the overlap of two orthogonal nematic domains. Therefore, domain boundaries provide a good platform to investigate the electronic properties of nematicity-absent states and the emergent phenomena associated with lattice strains, such as topological non-trivial states and superconductivity.[5-12] However, due to the complexity on both sides of sample preparation and precision measurement, the nematicity suppression on domain boundaries has not yet been reported.

In single crystals of FeSe, superconductivity coexists with nematicity at low temperature;[21, 22] while in multilayer FeSe film grown on strontium titanate (FeSe/STO), the enhanced nematicity induces stripe-type charge ordering and destroys superconductivity.[23] The nematic domain boundaries exist over large area of FeSe/STO,[23] which enables us to study their electronic properties. In our experiments, we grew ~ 20 UC FeSe films on STO substrates by using MBE method and investigated the electronic structures in FeSe films by STM.

The nematicity of FeSe film can be directly observed in STM topographic image with atomic resolution. Figure 1a shows the Se-terminated surface within a nematic domain. The image of every single Se atom is elongated along the diagonal direction of the Se-Se lattice (that is the direction of underlying Fe-Fe lattice), reflecting the electronic anisotropy (nematicity) in the Fe-plane. The capability to capture the signals of nematicity is realized when one atom or molecule sensitive to orbital component of FeSe is picked up to the STM tip apex. The case is also shown in Figure 1b where two adjacent FeSe nematic domains are separated by a domain boundary. As expected, the orientations of the elongated Se atoms in the two domains are perpendicular to each other. The most striking phenomenon is the recovery of the round shape of Se atoms on the domain boundary (see the schematic orange patches in Figure 1b), thus supporting the absence of nematicity in this area. The STM topographic image here directly records the image shape change of the Se atoms from the nematic region to the nematicity-absent one. The shape evolution occurs within the distance of $2a_{Se}$, indicating a rather sharp edge between the two regions. Therefore, FeSe film is composed of networks of nematicity-absent domain boundaries and the domains with strong nematicity.

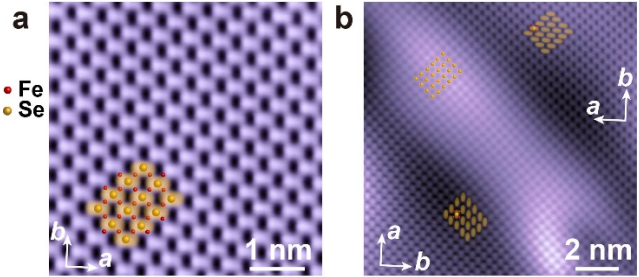

**Figure 1.** The absence of nematicity in the FeSe domain boundary. (a) The atomically resolved STM topographic image of a ~20 UC FeSe film grown on STO substrate (5 nm × 5 nm, set point: $V_s$ = 20 mV, $I_t$ = 0.2 nA). Se atoms are apparently elongated along the diagonal direction of the Se-Se lattice due to the large strength of nematicity in FeSe film. (b) The elongated Se atoms within the twin domains and the recovered round Se atoms on the boundary (13 nm × 13 nm, $V_s$ = 20 mV, $I_t$ = 0.2 nA). The orange patches indicate the apparent shapes of Se atoms at the corresponding locations.

The nematic domain boundaries are shown in Figure 2a and b. The boundaries acquired at -50 meV (Figure 2a) and 70 meV (Figure 2b) are of opposite contrasts. Meanwhile, the crossing points of several domain boundaries (indicated by the orange arrows) is dark in contrast at -50 meV and becomes bright at 70 meV. The crossing point corresponds to a dislocation and is related to the defects in the STO substrate. After careful inspection of large amount of domain walls, we find that the number of domain walls terminated at the crossing point is always four (see Figure 2c too), which is probably protected by the system symmetry.[24-26] Figure 2c shows a crossing point of four domain walls in detail, in which the traces of the boundaries are highlighted by the orange dashed lines. The stripes induced by impurities in each domain are visible[23] (the orientations of the stripes are denoted by the white double-headed arrows). The existence of the stripes reveals the strong nematicity in FeSe domains, and this nematicity is greatly suppressed on those domain boundaries.

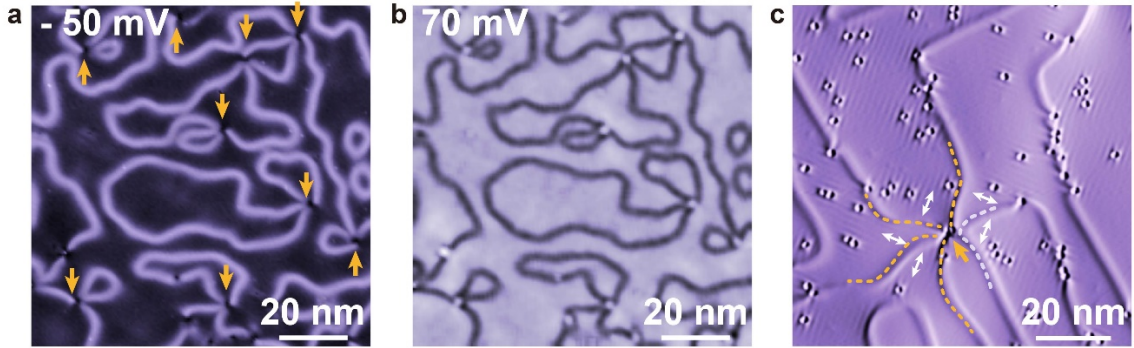

**Figure 2.** Nematic domain boundaries in ~20 UC FeSe/STO. (a and b) Spatially resolved density-of states maps of the nematic domain boundaries. (a) and (b) exhibit the domain walls with opposite contrasts at -50 meV and 70 meV, respectively (100 nm × 100 nm, $V_s$ = 250 mV, $I_t$ = 0.3 nA). The orange arrows indicate the crossing points of several domain walls. (c) STM topographic image of a FeSe film (90 nm × 90 nm, $V_s$ = 100 mV, $I_t$ = 50 pA). It shows a crossing point of four domain walls and the stripes within each domain. The traces of the domain walls in the vicinity of the crossing point are marked by the orange dashed lines. Note that the domain wall on the right side (marked with white dashed line) gets around rather than terminates at the crossing point, which is more clearly shown in Figure 4a and b. The stripes, originating from the large strength of nematicity, are pinned by impurities within the domains. The orientations of those stripes are denoted by the white double-headed arrows.

We now focus on the electronic structures of the domain boundaries. Scanning tunneling spectroscopy (STS, or d$I$/d$V$ spectrum) measures the local quasiparticle density of states (DOS) of the sample surface. Spatially resolved d$I$/d$V$ spectra are taken along two traces, across and along the center of a domain boundary in Figure 3a (see the dashed line Cut 1 and 2), respectively. Along the domain wall center (Cut 2), large-bias-range d$I$/d$V$ spectra (Figure 3c) exhibit high spatial homogeneity with a characteristic peak at -150 meV. Cut 1 crosses the two kinds of regions, the twin-domains and their boundary (domain wall), where the characteristic peak shifts downwards about 35 meV from -150 meV on the domain wall to -185 meV within the domain (Figure 3b). The energy shift of the peaks on the two regions may induce emergent states on the edge.

Small-bias-range d$I$/d$V$ spectra along Cut 1 in Figure 3d exhibit two distinct peaks at -10 meV and 0 meV. Superconductivity is absent in both regions, since no gap feature is observed near the Fermi level ($E_F$) in the spectra. The peak at -10 meV corresponds to the characteristic energy of the domain wall (See the spectra taken along Cut 2 in Figure 3e), and the DOS map at -10 meV (Figure 3f) sketches the contours of the domain wall. The DOS map at 0 meV ($E_F$, see in Figure 3g) reveals two striped structures, corresponding to the two induced edge states arising from the discrepancy of electronic structures between the two regions (see the orange arrows as reference in Figure 3a, f and g).

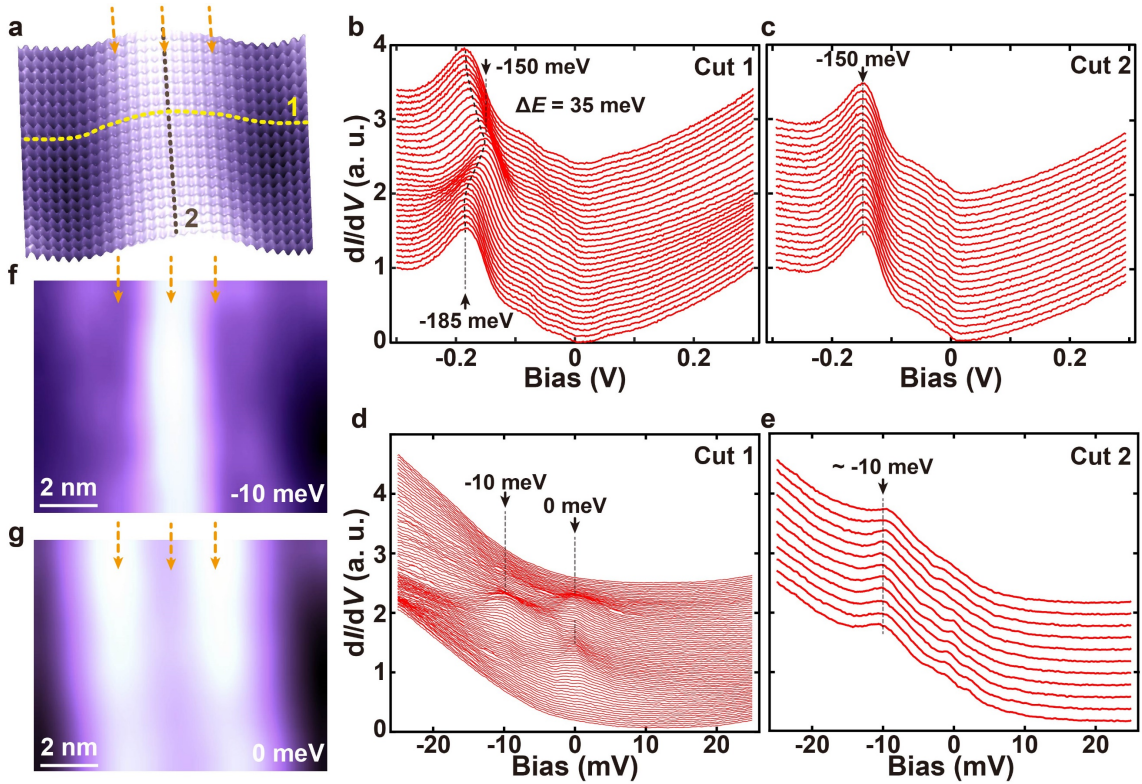

**Figure 3.** Edge state and electronic properties of the ~20 UC FeSe/STO domain boundaries. (a) Three-dimensional STM topographic image of a domain wall (10 nm × 8 nm, $V_s$ = 20 mV, $I_t$ = 0.2 nA). (b and c) Spatially resolved large-bias-range d$I$/d$V$ spectra taken along the dashed lines (Cut 1 and Cut 2) in (a). Spectra are shifted vertically for clarity. Set point: $V_s$ = -300 mV, $I_t$ = 0.2 nA. The characteristic peak shifts upwards ~ 35 meV on the domain wall. (d and e) Spatially resolved small-bias-range d$I$/d$V$ spectra taken along the dashed lines in (a), from which two peaks at 0 meV and -10 meV are revealed. Set point: $V_s$ = -25 mV, $I_t$ = 0.3 nA. (f and g) Spatially resolved DOS maps of (a) at -10 meV and 0 meV, respectively. Edge states are clearly shown in (g). Set point: $V_s$ = -25 mV, $I_t$ = 0.3 nA.

Interestingly, another bound state at $E_F$ is observed at the crossing point of the domain walls. Figure 4a and b show the DOS maps of a crossing point at $E_F$ and -10 meV, respectively. At $E_F$, an obviously higher DOS (denoted by the white arrow in Figure 4a) appears at the crossing point of four domain walls. The number four may be a topological invariant, as determined by the crystal symmetry of FeSe. The crossing point therefore is a possible topological defect.[24, 25] STS (Figure 4c) at the crossing point shows a pronounced peak near $E_F$, indicating a trapped bound state. The bound state may have different origin from that of the edge state.

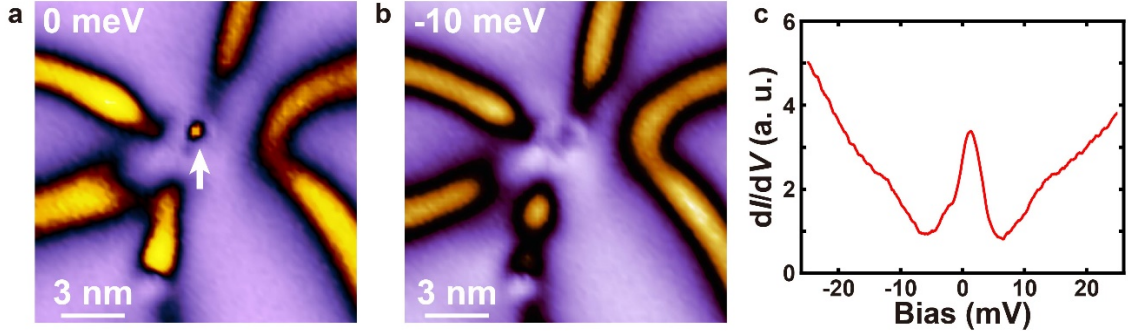

**Figure 4.** Bound state at the crossing point of the domain boundaries. (a and b) The DOS maps of the crossing point in Figure 1c (15 nm × 15 nm, $V_s$ = 100 mV, $I_t$ = 0.45 nA). A bound state around $E_F$ is trapped at the crossing point, marked by the white arrow. (c) d$I$/d$V$ spectrum of the crossing point. Set point: $V_s$ = -25 mV, $I_t$ = 0.3 nA.

We now discuss possible topological origin of the edge state. The discrepancy of the electronic structures on and off the domain boundaries mainly manifests as an energy shift of 35 meV in STS (Figure 3b). The characteristic peak at -185 meV originates from a rather flat band (corresponding to higher DOS) of FeSe around the Γ point below $E_F$, which is attributed to tunneling to Fe 3$d_{z^2}$ orbital ($d_{z^2}$ band).[3] The position of the $d_{z^2}$ band is not sensitive to nematicity, thus local lattice information, such as strain and anionic height, can be determined from the $d_{z^2}$ band energy. Bulk FeSe shows a peak at -220 meV,[27] corresponding to the original FeSe lattice without strain. Thickness dependent ARPES measurements on FeSe films indicate that the $d_{z^2}$ band moves upward with increasing tensile strain.[28] By comparison with ARPES results, the STS peak at -185 meV in the domain region corresponds to FeSe film with a thickness of 20 unit cells (UCs), consistent with the estimated value from the film growth rate. The position of the $d_{z^2}$ band at -150 meV at the domain wall region (Figure 3b and c), even shallower than that in 4 UC FeSe/STO,[28] indicates that its in-plane lattice constant is comparable to that of the STO substrate (0.39 nm). Such lattice expansion is confirmed by our density functional theory (DFT) calculations, in which the $d_{z^2}$ band at -150 meV corresponds to an in-plane lattice constant of 0.386 nm (Supporting Information, Section 1 and Figure S1). Therefore, comparing with that in the nematic domain, the in-plane lattice of the domain boundary is strongly elongated.

Topological non-trivial state is realized in such region with large lattice expansion through the band inversion at the M point, which is predicted by an analytic approach combined with symmetry analysis of FeSe film.[5] The mechanism relies on the intra-$d_{xy}$ and $d_{yz}$ bands crossing and the on-site spin-orbit coupling (SOC) between the electrons of $d_{xz}$ and $d_{yz}$ orbitals, leading to hybridized gap opening at the M point.[5,9] The hopping between $d_{xy}$ and $d_{yz}$ orbitals is sensitive to the in-plane lattice distortion. The in-plane lattice expansion as well as the decreasing Se height with respect to Fe-plane gives rise to enlarged bandwidth of $d$ orbital bands and hence promotes the band inversion at the M point. Regarding the band structure topology, at the M point the $d_{xz}$ and $d_{yz}$ bands are degenerate due to the absence of nematicity on the domain boundaries, which resembles the suppression of nematicity in monolayer FeSe/STO via carrier doping.[3,4] Therefore, the two key conditions for realizing topological phase,[5] tensile strain and absence of nematicity, are both well satisfied at the domain boundaries, making the observed edge and bound states a promising candidate for the topological non-trivial states.

The topological state discussed above is based on the absence of antiferromagnetism (AFM) in FeSe. However, the checkerboard AFM might develop in the FeSe domain boundaries, since the electronic properties, such as the broken two-fold symmetry and the in-plane lattice expansion, are similar to that in 1 UC FeSe /STO,[6] in which a recent study shows experimental evidence for AFM order.[29]

DFT calculations are carried out to study the possible topological origin of the edge states with AFM. For a two-dimensional (2D) film, the topological invariance usually oscillates with the number of layers[30,31]. Surprisingly, the edge states in even and odd layers of FeSe films exhibit similar behaviors in our STM measurements (Supporting Information, Section 3). Our theoretical analysis and DFT calculations provide a possible explanation for such intriguing phenomena (Figure 5 and Supporting Information, Section 2). One new type of topological crystalline insulating phase is found to guarantee the presence and robustness of the edge states on the boundary of even layers FeSe film. The edge states (Figure 5c) in odd FeSe layers are protected by a 2D antiferromagnetic $Z_2$ invariant,[32] related to the combined symmetry of time-reversal and primitive-lattice-translation; While the edge states (Figure 5d) in even layers are protected by the time-reversal related glide mirror symmetry. The distinction here resembles the important difference between 2D $Z_2$ topological insulator[33-37] and topological crystalline insulator.[38, 39]

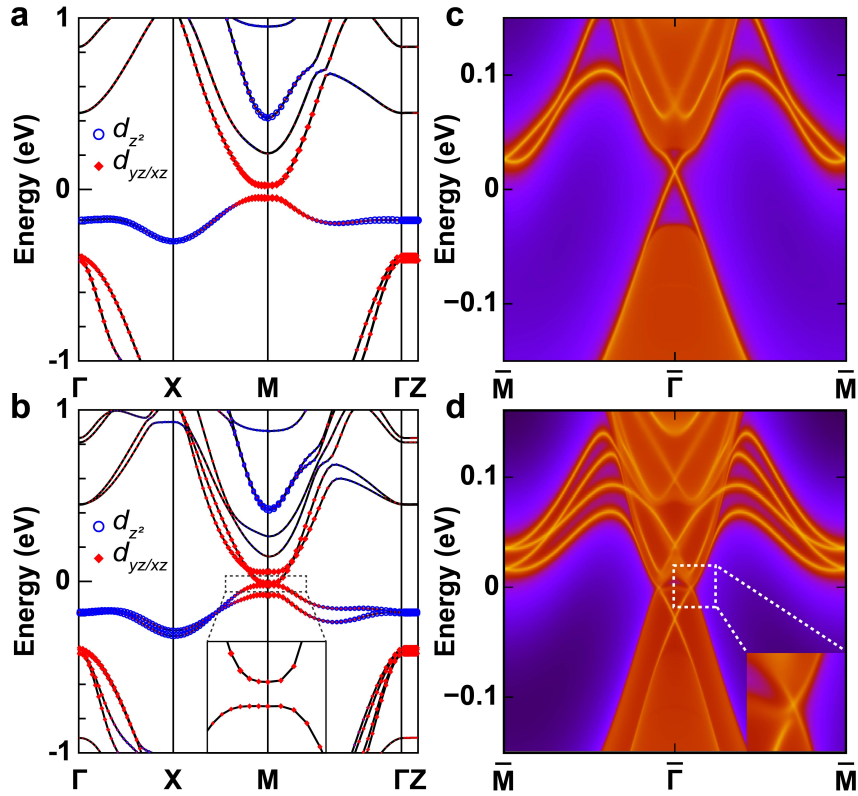

**Figure 5.** Band structures and topological edge states in odd and even FeSe layers. (a and b) The calculated band structures of monolayer and bilayer FeSe films, respectively. The red diamonds denote the projection of the $d_{yz}/d_{xz}$ orbitals of Fe atoms and the blue circles denote the $d_{z^2}$ orbital. (c and d) The calculated edge states along (110) direction on the boundary of monolayer and bilayer FeSe films, respectively. The edge state crossing at the $\bar{\Gamma}$ point in (c) is protected by the combined symmetry of time-reversal and primitive-lattice-translation, while the robustness of the edge state crossing in (d) is protected by the time-reversal related glide mirror symmetry. Although the origin for the two proposed types of topological edge states are different, it is hard to distinguish between them by STM

results. The increased thickness of FeSe layers may induce band dispersion along z-direction, giving rise to deceased gap sizes. However, the topological edge states can still be protected by the symmetry discussed above. More discussions are in Supporting Information, Section 2.

In addition, the characteristic energies of the edge state and bound state are both at $E_F$, indicating that the chemical potential here is located in the center of the hybridized gap. Therefore, introducing domain boundaries in a superconducting FeSe film can be a good approach to realizing topological superconductivity, and its advantage is that the complexity to combine several compounds is avoided so that all the topological features could be achieved in a single-component material.

**Sample Growth.** FeSe films were grown on Nb-doped (0.05% wt) $SrTiO_3$ (100). STO was degassed at 500 °C for several hours and subsequently annealed at 1200 °C for 20 min to obtain the atomic flat $TiO_2$-terminated surface. To prepare the FeSe film, high purity Fe (99.995%) and Se (99.9999%) were evaporated from two standard Knudsen cells and the substrates were kept at 380 °C. The growth was carried out under Se-rich condition with a nominal Se/Fe flux ratio of ~15. The as-grown FeSe films were annealed at 400 °C for several hours to remove the excess Se. The density of excess Se on the surface is checked by STM. Thickness of the films in this study are ~ 20 UC.

**STM**. *In-situ* STM experiments were conducted in an ultra-high vacuum (UHV) low temperature (0.4 K) STM equipped with a MBE chamber for sample growth (Unisoku). A polycrystalline PtIr STM tip was used and characterized on Ag island before STM experiments.

**Calculations.** The DFT calculations were carried out by using the full-potential linearized-augmented plane-wave (FP-LAPW) method implemented in WIEN2K package.[40] Perdew-Burke-Ernzerhof (PBE)[41] type of generalized gradient approximation + Hubbard U (GGA+U) approach[42] with U = 1.0 eV (J = 0.2 eV) on the Fe's *3d* orbitals was used to treat the exchange and correlation potential. Spin-orbit coupling (SOC) was included as a second variational step self-consistently. The radii of the muffin-tin spheres ($R_{MT}$) were chosen as 2.27 a.u. and 2.16 a.u. for Fe and Se, respectively. The cutoff of the wave functions $K_{max} \times R_{MT} = 7$, and 15×15×2 k-point mesh was used in the self-consistent calculations. Maximally localized Wannier functions (MLWFs)[43] for the *3d* orbitals of Fe and the *4p* orbitals of Se were generated to construct the tight-binding (TB) Hamiltonians of semi-infinite sample and calculate the edge states iteratively.[44, 45]

**Data Availability**. Data will be available from the corresponding authors on request.

## ASSOCIATED CONTENT

**Supporting Information**

1. The evolution of the $d_z^2$ band with increasing in-plane lattice constant of FeSe.

2. Topological edge states in odd and even FeSe layers.

3. Experimental evidence of edge states in odd and even FeSe layers.

## AUTHOR INFORMATION

**Corresponding Authors**
*Wei Li Email: weili83@tsinghua.edu.cn Phone: +8610-62795838
*Gang Xu Email: gangxu@hust.edu.cn Phone: +8627-87792334-8120
*Qi-Kun Xue Email: qkxue@mail.tsinghua.edu.cn Phone: +8610-62795618

**Author Contributions**
W.L. and Q-K.X. designed and coordinated the experiments; Y.Y., W.L., P.D. and Z.X. carried out the STM experiments; Y.Y., and W.L. grew the samples; B.L. and G.X. performed theoretical calculations. X.C., X.M., K.H., L.W. and C.S. provided discussion about the data. W.L. wrote the manuscript with comments from all authors.

**Notes**
The authors declare no competing financial interest.

## ACKNOWLEDGEMENTS

We thank Y. Zhang, J. P. Hu, H. Yao and Z. Liu for helpful discussions. The experimental work was supported by the National Science Foundation (No. 11674191), Ministry of Science and Technology of China (No. 2016YFA0301002) and the Beijing Advanced Innovation Center for Future Chip (ICFC). W. Li was also supported by Tsinghua University Initiative Scientific Research Program, Beijing Young Talents Plan and the National Thousand-Young-Talents Program.